\documentclass[12pt]{article}
\usepackage{fullpage,graphicx,psfrag,amsmath,amsfonts,verbatim,cleveref}
\usepackage[small,bf]{caption}
\usepackage{subcaption}
\usepackage{url}
\usepackage{booktabs}

\newcommand{\ones}{\mathbf 1}
\newcommand{\reals}{{\mbox{\bf R}}}

\newcommand{\diag}{\mathop{\bf diag}}

\newcommand{\dom}{\mathop{\bf dom}} 

\newcommand{\eg}{{\it e.g.}}
\newcommand{\ie}{{\it i.e.}}

\title{Driver Positioning and Incentive Budgeting with an Escrow Mechanism for Ridesharing Platforms}
\author{Hao Yi Ong \and Daniel Freund \and Davide Crapis}

\begin{document}
\maketitle

\begin{abstract}
Drivers on the Lyft rideshare platform do not always know where the areas of supply shortage are in real time. This lack of information hurts both riders trying to find a ride and drivers trying to determine how to maximize their earnings opportunity. Lyft’s Personal Power Zone (PPZ) product helps the company to maintain high levels of service on the platform by influencing the spatial distribution of drivers in real time via monetary incentives that encourage them to reposition their vehicles. The underlying system that powers the product has two main components: (1) a novel “escrow mechanism” that tracks available incentive budgets tied to locations within a city in real time, and (2) an algorithm that solves the stochastic driver positioning problem to maximize short-run revenue from riders’ fares. The optimization problem is a multiagent dynamic program that is too complicated to solve optimally for our large-scale application. Our approach is to decompose it into two subproblems. The first determines the set of drivers to incentivize and where to incentivize them to position themselves. The second determines how to fund each incentive using the escrow budget. By formulating it as two convex programs, we are able to use commercial solvers that find the optimal solution in a matter of seconds. Rolled out to all 320 cities in which Lyft’s operates in a little over a year, the system now generates millions of bonuses that incentivize hundreds of thousands of active drivers to optimally position themselves in anticipation of ride requests every week. Together, the PPZ product and its underlying algorithms represent a paradigm shift in how Lyft drivers drive and generate earnings on the platform. Its direct business impact has been a 0.5\% increase in incremental bookings, amounting to tens of millions of dollars per year. In addition, the product has brought about significant improvements to the driver and rider experience on the platform. These include statistically significant reductions in pick-up times and ride cancellations. Finally, internal surveys reveal that the vast majority of drivers prefer PPZs over the legacy system.
\end{abstract}

\newpage
\tableofcontents
\newpage

\section{Introduction}

\begin{figure}[t]
    \centering
    \begin{subfigure}{.5\linewidth}
        \centering
        \includegraphics[width=0.5\linewidth]{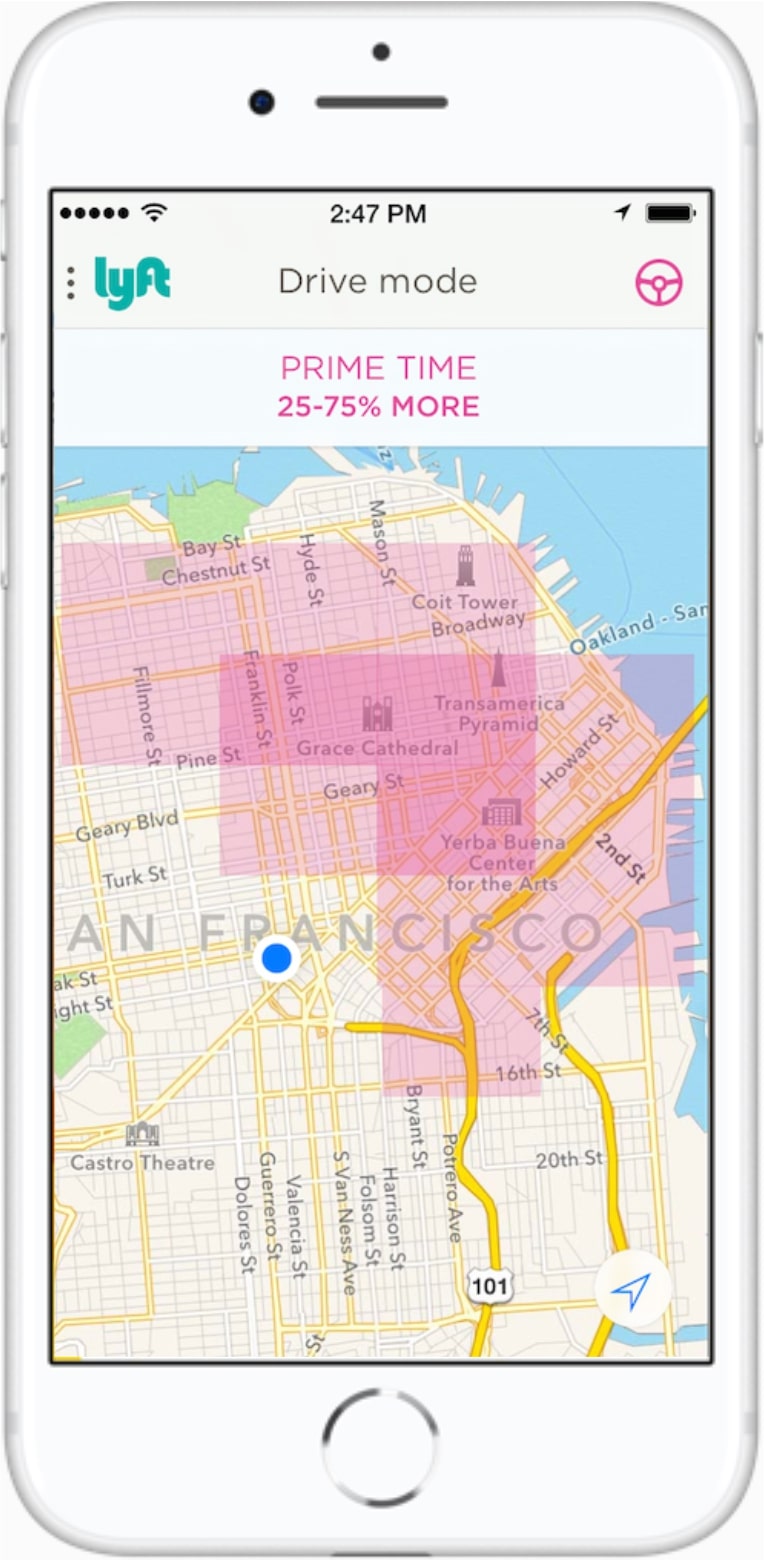}
    \end{subfigure}%
    \begin{subfigure}{.5\linewidth}
        \centering
        \includegraphics[width=0.5\linewidth]{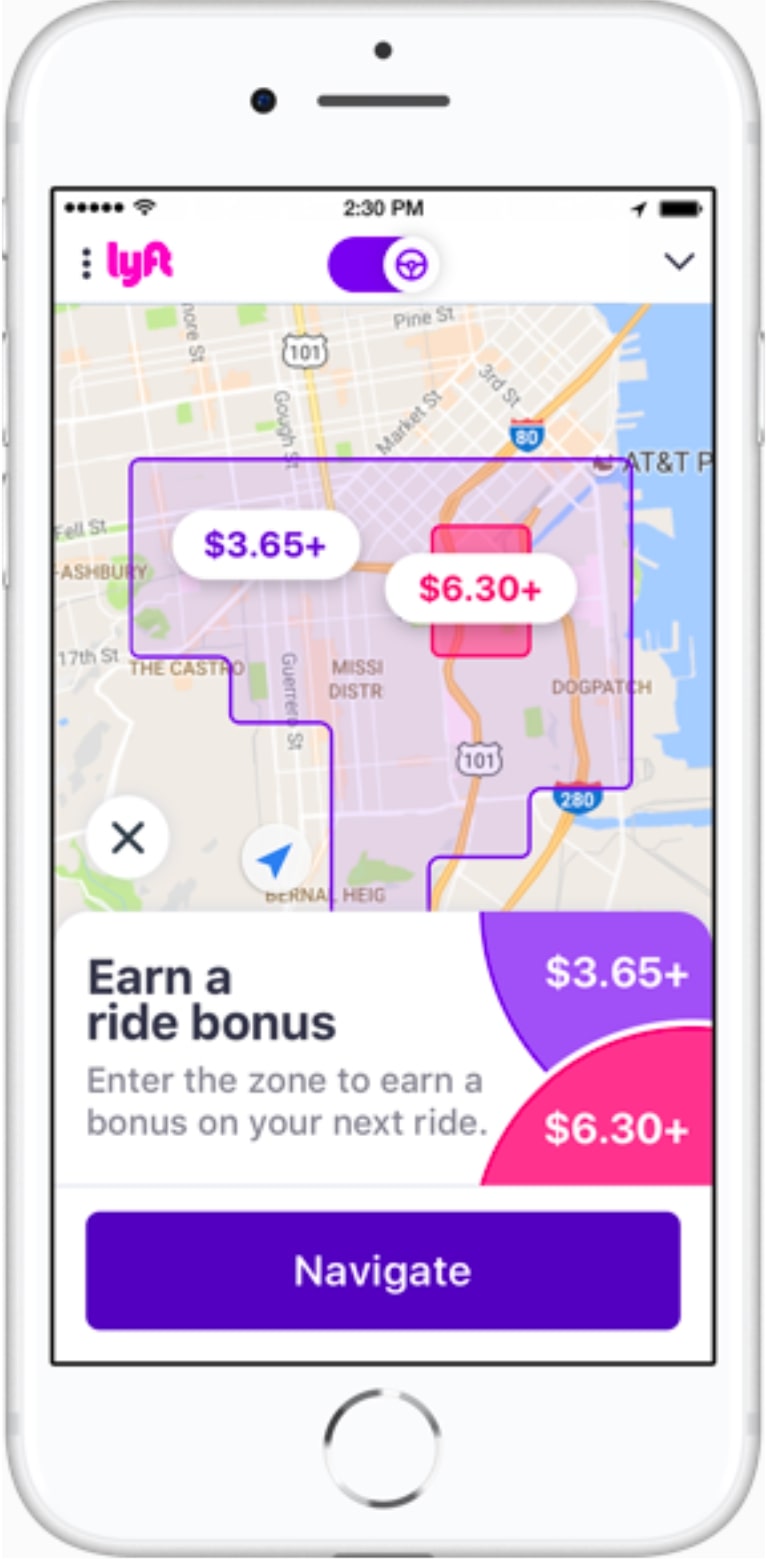}
    \end{subfigure}
    \caption{PT adds a predetermined percentage of the base fare onto the driver’s total fare (left), whereas drivers accrue a Personal Power Zone bonus when they enter the corresponding zones (right).}
    \label{Figure1}
\end{figure}

Lyft, Inc. develops and operates a mobile application (app), offering a ridesharing platform, as well as motorized scooter and bicycle sharing services. The company is based in San Francisco, California and operates in 320 cities in the United States and Canada. The biggest and most mature part of its business is ride sharing, a marketplace in which riders are matched in real time to drivers who drive them to their destinations. Our work seeks to incentivize the improved spatial positioning of drivers, so that drivers might better meet rider demand. The work was conducted with Lyft’s Driver Positioning team, a cross-functional team of product managers and designers, software engineers, and research scientists that work as a unit to tackle problems related to the supply side of the market and develop products to improve this side of the marketplace.
In this paper, we describe the development of Lyft’s Personalized Power Zone (PPZ) product. PPZ is an innovation that Lyft introduced to improve the drivers’ experiences during peak demand periods. Historically, ride-hailing platforms have focused on dynamic pricing to match rider demand to available driver supply, charging higher prices during periods of heightened demand. Lyft calls its dynamic-pricing product “Prime Time” (PT) and Uber calls its equivalent “Surge.” These are multiplicative modifiers on top of the base time-and-distance fare, thereby reactively suppressing rider demand through marked-up fares. For drivers, both platforms have traditionally used heatmaps, which show the magnitude of elevated ride fares as colors on city maps as a visual aid to indicate locations with elevated demand; see the left subfigure in \Cref{Figure1}. Since the driver’s PT bonus in the legacy system was proportional to the ride’s base fare and the PT multiplier, heatmaps serve as an incentive that drivers can use to reposition themselves toward areas with high PT multipliers \cite{LFK18}. 

\begin{figure}[t]
    \centering
    \includegraphics[width=0.6\linewidth]{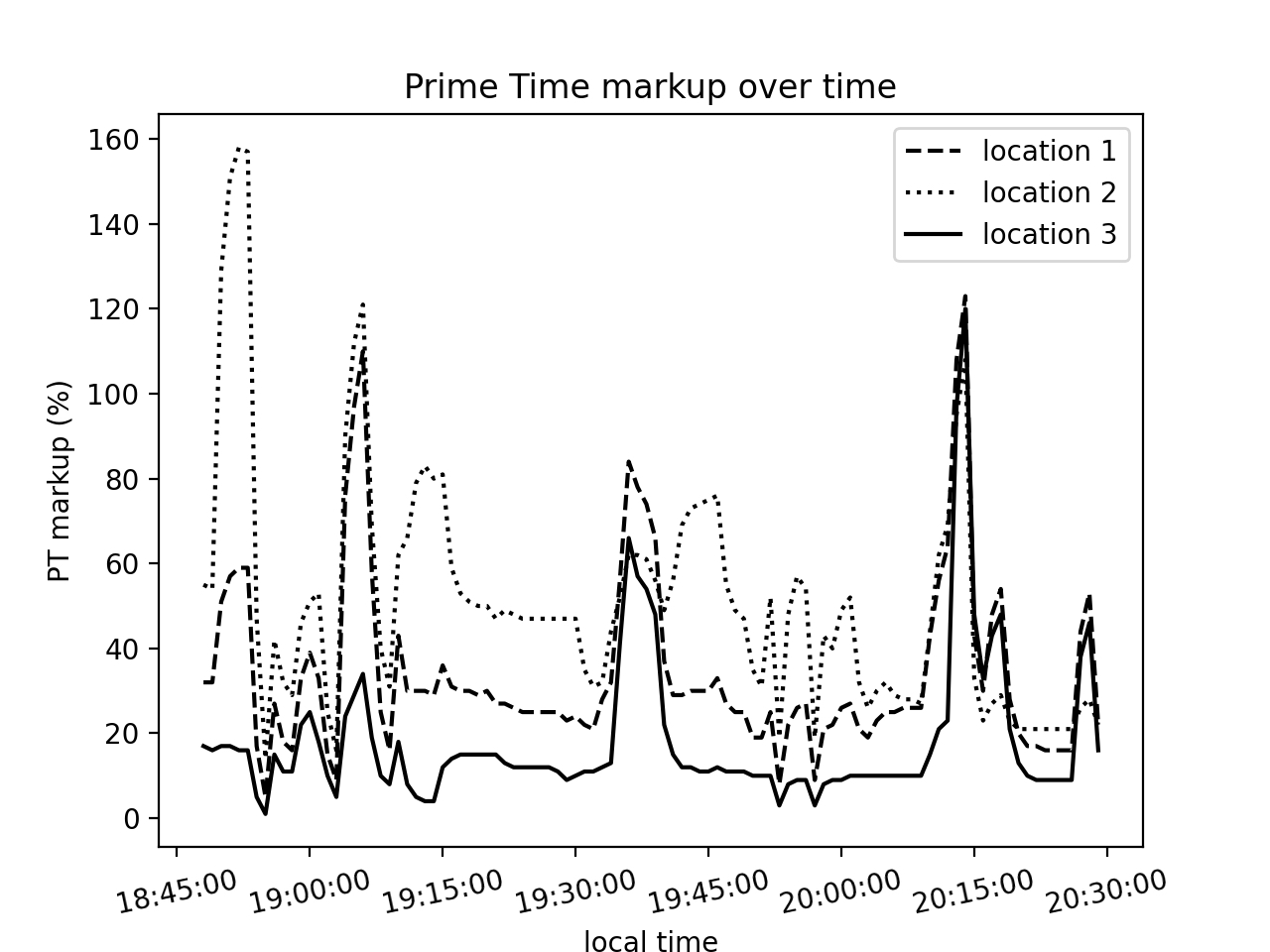}
    \caption{
    PT bonus multiplier values plotted for three nearby locations illustrate how spatially and temporally volatile they can be over a short time span.
    }
    \label{Figure2}
\end{figure}

Generally, dynamic pricing provides a high earning potential for drivers, especially around persistent heatmap “hotspots,” which are zones on the city map with elevated prices. In a commission-based platform system, drivers benefit from the rides’ higher prices (i.e., higher \$/minute spent driving a passenger). In addition, greater demand can give rise to greater driver utilization (i.e., minutes spent driving passengers/total minutes driving), as long as the time drivers spend per ride does not increase due to increased pick-up times, often referred to as estimated time of arrival (ETA) \cite{CKW17}. Nonetheless, prior to the implementation of PPZs, the driver experience in peak demand periods was never optimized for drivers. Although some evidence exists showing that drivers react to the heatmap \cite{LFK18}, the predominant wisdom among drivers has been that actively chasing heatmap hotspots is not a good strategy to maximize earnings \cite{Gri17}. In large part, this is motivated by the unpredictable and fast-paced updates of PT levels set by dynamic pricing algorithms that adapt rapidly to changes in the marketplace (see \Cref{Figure2}). The goal of PPZs was to replace the driver’s PT heatmap and adapt the drivers’ compensation during PT in a manner that simultaneously incentivizes improved driver positioning within a city and the drivers’ experience by more directly rewarding the positioning effort.

Replacing the driver PT product required a significant shift to Lyft’s ride-hailing platform design. For example, it involved deviating from a fixed in-ride, on-trip commission-based model that pays the driver a constant fraction of the rider’s PT surcharge. This change was necessary to improve upon the shortcomings of the driver PT experience, which we outline in additional detail below. However, beyond this design change, it also required significant algorithmic innovations. Our paper highlights these algorithmic innovations and the results they produced. The two key algorithmic developments that we describe are as follows.

\begin{itemize}
    \item A robust spatial budget-tracking mechanism that provides real-time demand signals, which we discuss in the \emph{Budgeting Via a Location-Based Escrow Mechanism} subsection.
    \item An asymptotically optimal two-stage algorithm that produces driver relocation incentives to maximize market efficiency, which we describe in the \emph{Problem Decomposition and Certainty Equivalent Approximation} subsection.
\end{itemize}

Beyond the change to Lyft’s platform, the PPZ development also deviated significantly from the ideas traditionally considered in the academic study of the gig economy. In these works, the focus has usually been on the platform taking a constant proportion of each trip’s fare, including the PT portion. Only recently have Garg and Nazerzadeh \cite{GN21} considered a stylized model that exposes fundamental limitations that arise with a multiplicative PT bonus due to the fast-paced changes in the levels of dynamic rider prices often observed in the market. (Note that Garg and Nazerzadeh was originally written in 2019 and was updated in 2021.) To the best of our knowledge, all prior work had either explicitly (e.g., \cite{banerjee2015pricing}, \cite{bimpikis2016spatial}, \cite{cachon2017role}) or implicitly (e.g., \cite{ma2018spatio} considered models in which the drivers’ bonus was a constant proportion of the rider’s PT. In contrast to Garg and Nazerzadeh \cite{GN21}, we describe a real-world implementation of a system that considers this concern and others, which we describe in the \emph{Problem Background} section.

We structured the remainder of this paper as follows. In the \emph{Problem Background} and \emph{Product Description} sections, we discuss the shortcomings of the legacy system and the PPZ product design. We describe the key technical challenges to overcome when implementing PPZs in \emph{Technical Implementation Challenges}. In the \emph{Stochastic Model} and \emph{Optimization Approach} sections, respectively, we describe the theoretical model that motivated our algorithmic approach and the approach. In \emph{Implementation}, \emph{Numerical Experiments}, and \emph{Live Experiments}, we outline the challenges that arose in measuring the business impact of the new system and the causal inference approach we developed to overcome them. We conclude with a discussion of the impacts these innovations enabled in the \emph{Broader Impact} section. In the appendices, we provide a formal description of the technical challenge and our optimization model. The algorithms we present are currently deployed at Lyft in the 320 cities in which we operate. Their cumulative impact has included an increase in yearly bookings by tens of millions of dollars and a reduction in driver cancellation rates by 13\%. Every week, these algorithms generate millions of PPZs that help hundreds of thousands active drivers decide where to drive.

\section{Problem Background}

We begin by discussing the limitations of the legacy driver PT system. The papers most similar to ours focus on the problem of setting prices to incentivize drivers to relocate toward high-earning opportunities. The key component the literature misses, to the best of our knowledge, is the significant and unpredictable spatial and temporal volatility of rider PT. To keep supply and demand in balance in a fast-paced market such as ride-hailing, the platform frequently updates its prices; for example. Uber updates its prices every two minutes \cite{LFK18}. Consider the time series plot of the PT bonus multiplier levels in three busy nearby locations (i.e., within a mile apart) in the greater San Francisco Bay Area, as we show in \Cref{Figure2}. Although the price-setting algorithm enforces some level of spatiotemporal smoothness, market conditions and thus PT bonus multiplier values change rapidly across time and space. In addition, the direction and magnitude of the changes are unpredictable. This holds true for platform designers with full information, and even more so for drivers with less information. Thus, common assumptions around drivers being rational agents with full information fail to capture the uncertainty inherent in the driver’s decision-making process. Below, we describe these complications and how we designed PPZs to overcome them.

\subsection{Cherry-picking of Long/Short Rides}

Garg and Nazerzadeh \cite{GN21} show that drivers may cherry-pick long rides during PT periods and short rides during non-PT periods to maximize their earnings using the legacy system. They consider a two-state Markovian system in which the state characterizes whether the current rider price is either low or high (i.e., the dynamic pricing is abstracted to being only a binary state). A driver receives a stream of ride requests of different trip lengths; the state of the system determines the rate at which the driver receives requests and the pay per minute and per mile of driving. The driver’s policy is to pick the set of requests to accept in each state. The main insight of Garg and Nazerzadeh is that by making the bonus for PT rides additive, rather than proportional to the trip length, the platform induces the driver to accept a larger set of trips. Intuitively, this is because accepting a short trip during surge periods may provide only a small earnings boost if the bonus is proportional to the (short) trip; yet, at the end of the trip, the surge period with its elevated opportunities may be over. Furthermore, they show (in their Theorem 3) that a bonus that is an affine function of the trip length is the best the platform can do to incentivize drivers to accept as many trips as possible. Although this constitutes a significant conceptual contribution toward an understanding of the fundamental limitations that arise under the legacy model, their stylized model does not give rise to a practical algorithm to overcome these shortcomings. Beyond implementing a product that capitalizes on their main insights, PPZ also addresses the following concerns that are not captured in the Garg and Nazerzadeh \cite{GN21} model.

\subsection{Spatiotemporal Volatility}

As we show in \Cref{Figure2}, PT is not only volatile in time but also in space. In particular, because different regions within a city experience different supply-demand imbalances, the platform prices rides differently across locations. Such spatial differences in prices are widely captured in the literature; examples include Bimpikis et al. \cite{bimpikis2016spatial}, Afeche et al. \cite{afeche2018ride}, and Ma et al. \cite{ma2018spatio}. However, the literature does not account for the interplay between spatial and temporal volatility. Specifically, between the time a driver sees prices induced by the supply-demand imbalance in a different location and the time that driver arrives at such a location after repositioning, the balance between supply and demand in the market may have changed substantially and the opportunity to boost his/her earnings may have disappeared.
This experience, common among drivers, has led experienced drivers to recommend not
“chasing the surge” \cite{Gri17}. In contrast, Bimpikis et al. \cite{bimpikis2016spatial}, Afeche et al. \cite{afeche2018ride}, Castro et al. \cite{castro2018surge}, and Ma et al. \cite{ma2018spatio} and many others assume that the drivers know the exact earnings opportunity arising from repositioning to a new location.

\subsection{Spatial Volatility}

Above we illustrated how temporal volatility discourages drivers from spatially repositioning. However, spatial price volatility also affects drivers who are already in an area where prices are high. Recall that under the legacy system, drivers earn a direct cut of the fare paid by the rider. In that world, a driver may encounter situations in which he/she is idling in a high-PT location and yet is dispatched to a nearby location with lower (or even no) PT. Ironically, the high PT at the driver’s location may dissuade a customer in that location from requesting a ride. Thus, the request-suppressing effect of PT may cause the driver to experience a smaller PT bonus and a longer ETA to serve the ride.

\subsection{Coordination}

Finally, many papers in the ridesharing context, including Ma et al. \cite{ma2018spatio}, Castro et al. \cite{castro2018surge}, Afeche et al. \cite{afeche2018ride}, Yang et al. \cite{yang2018mean}, and Bimpikis et al. \cite{bimpikis2016spatial}, assume different forms of spatial equilibria arising from drivers selfishly optimizing their own earnings. In practice, it is difficult to imagine how such equilibria would emerge in a dynamic system without agents having the ability to observe the actions of other drivers. Consider a group of drivers trying to reposition toward two different locations with PT. Even if they coordinated on the number of drivers to reposition to each location, they would need to make this decision based only on the observed PT levels. Because these would roughly correspond to relative supply shortfall, and not to absolute supply shortfall (see \emph{Price modifiers and elasticity} in \emph{Appendix A: Optimization Approach}), successful coordination is extremely unlikely.

\subsection{Limitations in the Legacy System}

In summary, despite providing drivers with higher earnings opportunities, the legacy system had misalignments between (1) the drivers' earnings maximization behaviors, (2) the behaviors the platform incentivizes (e.g., repositioning toward PT), and (3) the driver behaviors that would maximize platform metrics, such as number of rides served. Cumulatively, these led to both a poorer driver experience and platform inefficiencies. Noticeably, these issues could not have been addressed through mere algorithmic improvements to dynamic pricing. For example, they could not be resolved through a better forecast of riders' willingness to pay, or a change in the optimization problem underlying the price-setting process. Instead, a more fundamental design change was required to improve the driver experience during peak demand periods.

\section{Product Description}

The goal of the PPZ project was to optimize the PT experience for drivers by channeling the riders’ PT payments to drivers in a way that would resolve the limitations above and thereby improve the platform’s market metrics (e.g., bookings and driver hours); see the \emph{Live Experiments: Results} subsection below. From the driver’s perspective, a PPZ incentive consists of a delimited geographic area associated with a fixed, visible bonus. The driver qualifies for the bonus when entering the area and receives it upon completing his/her next ride request; the right side of \Cref{Figure1} shows the PPZ driver interface, including the bonus the driver accrues upon entering the purple/pink PPZ zone. By providing a clear reward for an unambiguous action, PPZ removes the uncertainty in the legacy system and thus creates an effective nudge for drivers to reposition. Beyond removing the issues surrounding uncertainty, globally optimizing the incentives given to drivers across a city also allows the PPZ product to globally coordinate the drivers’ locations. In addition, as PPZ replaces the proportional PT bonus of the legacy system, the driver may also receive an after-ride adjustment to ensure that the gap between the rider’s fare and the driver’s earnings is not excessively large. The additive upfront bonus experience, in conjunction with such after-ride adjustments based on time and distance, are similar to the affine bonuses suggested by Garg and Nazerzadeh \cite{GN21} and thus partially address the issues of drivers strategically rejecting trips to boost their earnings or cherry-picking their rides.

\section{Implementation Challenges}

A successful implementation of PPZs had to address two key technical challenges. First, PPZ would need to incentivize a better positioning of drivers than the legacy system. Second, the total amount of money spent on the PPZ incentives should match its budget (i.e., the PT paid by riders). In the following subsections, we describe these challenges and our initial attempts to overcome them.

\subsection{Repositioning}

Within any city in which Lyft operates, a set of drivers, referred to as open drivers or open supply, is available for dispatch. PPZs encourage open drivers to reposition to a different location by offering a monetary incentive at that location. The previous approach to incentive-based repositioning was based on the marginal value of supply in all spatial units of the city. Value can refer to rider request conversions or to the revenue they represent. This concept, which we refer to as the \emph{local sensitivity approach}, was to incentivize drivers to reposition from one location to another if a significant differential in the marginal value of supply between the two locations existed.

\begin{figure}[t]
    \centering
    \includegraphics[width=0.6\linewidth]{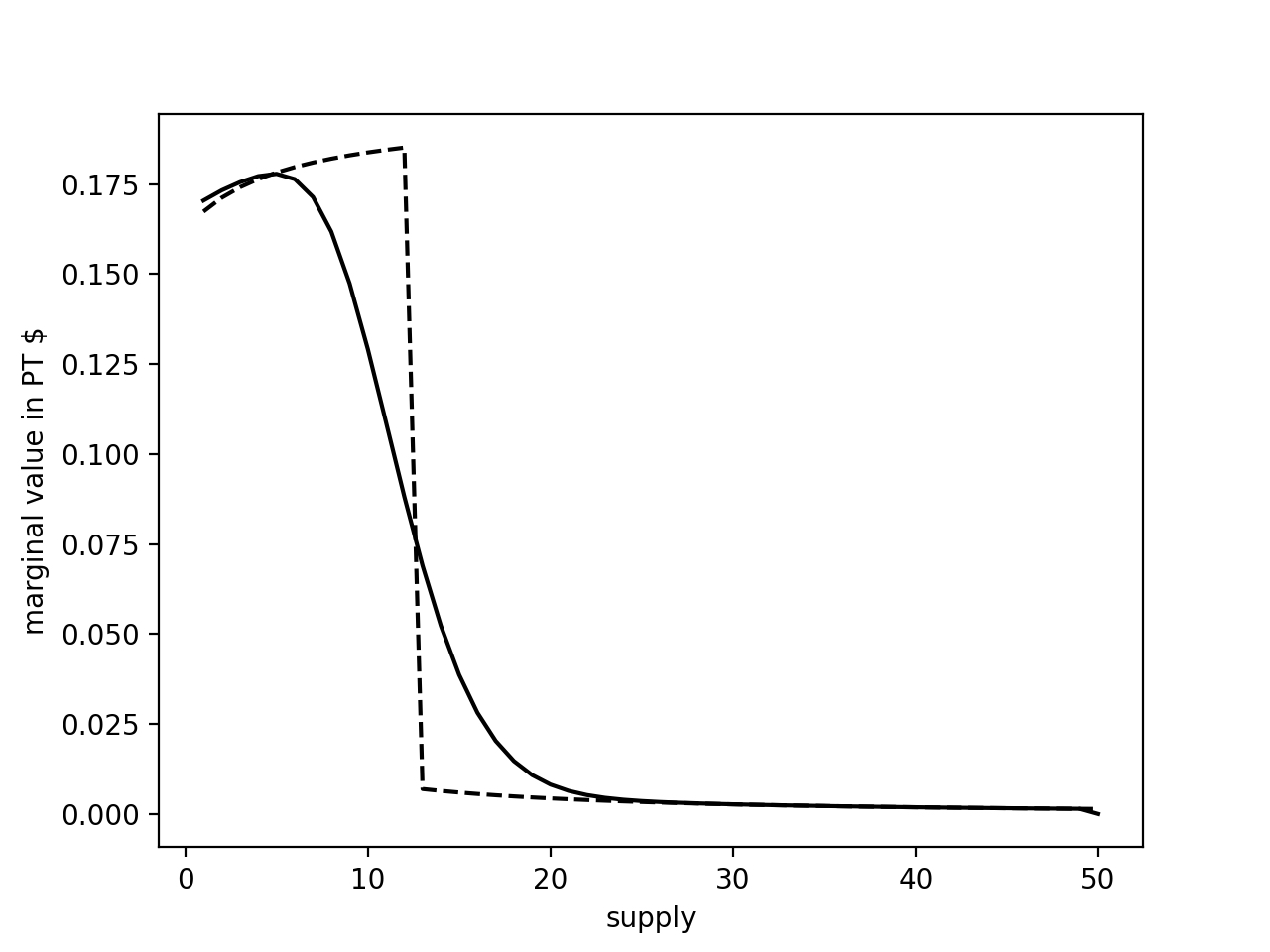}
    \caption{
    The graph shows the marginal value of supply evaluated at some location, as the highly nonlinear, step function-like empirically-fitted (solid) and theoretical (dashed) curves illustrate.
    }
    \label{Figure3}
\end{figure}

Unfortunately, the local nature of the approach did not account for the highly interactive nature of incentivizing a large number of drivers to reposition within a city. Implicitly, the local-sensitivity approach assumes that the local approximation (i.e., the gradient of our value function with respect to supply) provides a good indication of the market state even when it is perturbed by incentivizing many drivers to move simultaneously. This was a poor assumption. In practice, we observed that although the gradient evaluated at the current market state is accurate in the case of a small change in supply, incentivizing a large number of drivers to move breaks the locality assumption. The marginal value of supply is highly nonlinear and drops steeply beyond a specified supply count. \Cref{Figure3} shows an example of a location at which the nonlinearity is exhibited in both the empirically observed (solid line) and theoretically computed marginal values (dashed line). The steep drop in value appears when the supply count matches the number of ride requests at the location. In \emph{Appendix B: Perturbation and Sensitivity Analysis}, we prove that not only is the local sensitivity analysis largely invalid, but the decisions also made based on this approach can hurt market conditions.

Because PT is based on a multiplicative price markup on top of the eventual time-and-distance-based trip fare, the source of our budget is dynamic and is only fully realized after a ride has been completed. Specifically, we cannot know the value of the ride fare that funds the PPZ bonus until much later when the ride has been completed, or at least requested. To encourage a driver to reposition toward better ride opportunities, we must offer a PPZ bonus before the ride, which is intended to fund the PPZ bonus. This chicken-or-egg problem was at the source of PPZ’s budget-control challenge.

In the initial marginal-value-of-supply approach, the bonuses were paid as an affine function of the marginal value. Unfortunately, these bonus values often grossly overestimated the available funds, leading to overspending of up to 15 times our budget at times. This was in large part because the marginal value is a poor measure of the incremental revenue we can derive from the market. The spending problem became so severe that market degradation forced us to roll back the PPZ product from a major city in which we had already launched it \cite{ridesharePPZpause}. If we were to eventually scale the product to all 320 cities, we had to develop a much more robust way to control spending.

\section{Stochastic Model}

In this section, we describe a stochastic model that motivated our eventual optimization approach. The model is based on a multistage stochastic process on a discrete network of locations. Below, we outline the steps in each stage. We begin by describing what is assumed to be known at the beginning of the first stage. Initially, we know for each location~$i$ the current number of idle drivers and the expected demand. In addition, we know for any two locations~$i$ and~$j$ (1) whether we can dispatch a driver from~$i$ to serve demand in~$j$, and (2) the \emph{response probability} of a driver in~$i$ to relocate to~$j$ (i.e., the probability that a driver will relocate from~$i$ to~$j$ when a PPZ incentivizes that driver to do so). Finally, for each location we have a current budget estimate. We provide details in the \emph{Budgeting Via a Location-Based Escrow Mechanism subsection} below.

\subsection{Supply}

Based on the above information, we must determine the set of drivers to whom we will provide an incentive to relocate, and the destination to which we want these drivers to relocate. In doing so, we are constrained to not overspend the budget. Each driver who is given an incentive makes a stochastic decision whether to reposition based on that driver’s present location and the incentivized destination.

\subsection{Demand}

After the drivers relocate, the platform sets the PT price markups based on the new driver locations and the forecast demand. The prices are set as part of an optimization problem to maximize a metric, which we refer to as \emph{no-PT bookings}. This metric captures the time-and-distance fare of all serviced ride requests but does not include the PT markups. The no-PT bookings metric is one of Lyft’s most important metrics because it captures revenue without rewarding markets with mismatched supply and demand. For example, when including PT, bookings may be higher in the short run due to extreme PT markups; however, these are often perceived as detrimental to rider retention in the long run. Thus, no-PT bookings is the main metric we aim to maximize with PPZs. The constraints of the underlying optimization problem ensure that the expected number of ride requests, suppressed by the PT price markups, does not deplete the supply beyond a \emph{reserve level} that ensures that ETAs remain acceptably low \cite{CKW17}. After the platform sets its prices, we model the arrival of ride requests as a Poisson process with a price-dependent rate. Based upon the realized ride requests and the incentive-repositioned supply, the platform dispatches drivers to riders.

\section{Optimization Approach}

In this section, we provide a high-level overview of the methods we use to solve the problem described in the \emph{Stochastic Model} section above. As part of the solution, we also describe a novel \emph{escrow mechanism} used to generate budgetary signals for our algorithm. We provide a technically rigorous and detailed discussion of our approach in \emph{Appendix A: Optimization Approach}.

\subsection{Budgeting Via a Location-based Escrow Mechanism}

Our second key challenge was to spend a highly dynamic budget before we collect it. Recall the budget materializes only upon ride completion via rider PT payments. This dynamic budget is also highly location dependent because demand levels can vary significantly across a city. Given the delay between the time at which we must spend and the time at which the budget is realized in our accounts, we would like to accurately predict \emph{when} and \emph{where} the budget income may materialize; however, due to the spatiotemporal volatility of PT, doing so is impossible. To circumvent this conundrum, our key insight was that fast, real-time data based on upfront fares at ride dispatch can substitute for good predictions. Intuitively, we treat the PT paid by riders as part of our budget at the time and place of the ride request and dispatch despite the risk of the ride still being canceled or the PT amount changing because the rider changes the origin or destination. Upon completion of the trip, we then correct for the difference in amount that was realized. We note that although ride requests are a lagging indicator for future demand, we know empirically that rider demand (and the elasticity of demand) is generally well-correlated in the same location within a reasonably short time duration. Thus, as long as we can spend the budget accumulating from upfront fares quickly, the location of incoming PT across the city at any point in time will provide a good indication of where we should provide PPZs.

To track how much to spend and therefore how much money to offer per incentive, we create a set of virtual accounts associated with the locations from which ride requests originate. We call these virtual accounts a city’s ``local accounts.'' The escrow mechanism updates in response to events such as when a driver receives a PPZ or accepts a ride request and when a rider cancels a ride request or is dropped off at the ride destination. The occurrence of these events correspond to when and where we can account for the estimated and actual financial transactions, which determine the portions of the PT income earmarked for PPZ spend and the PPZ spend itself. As these events occur, we actively update the balance of the location-based escrow accounts in real time.

\begin{figure}[t]
    \centering
    \includegraphics[width=\linewidth]{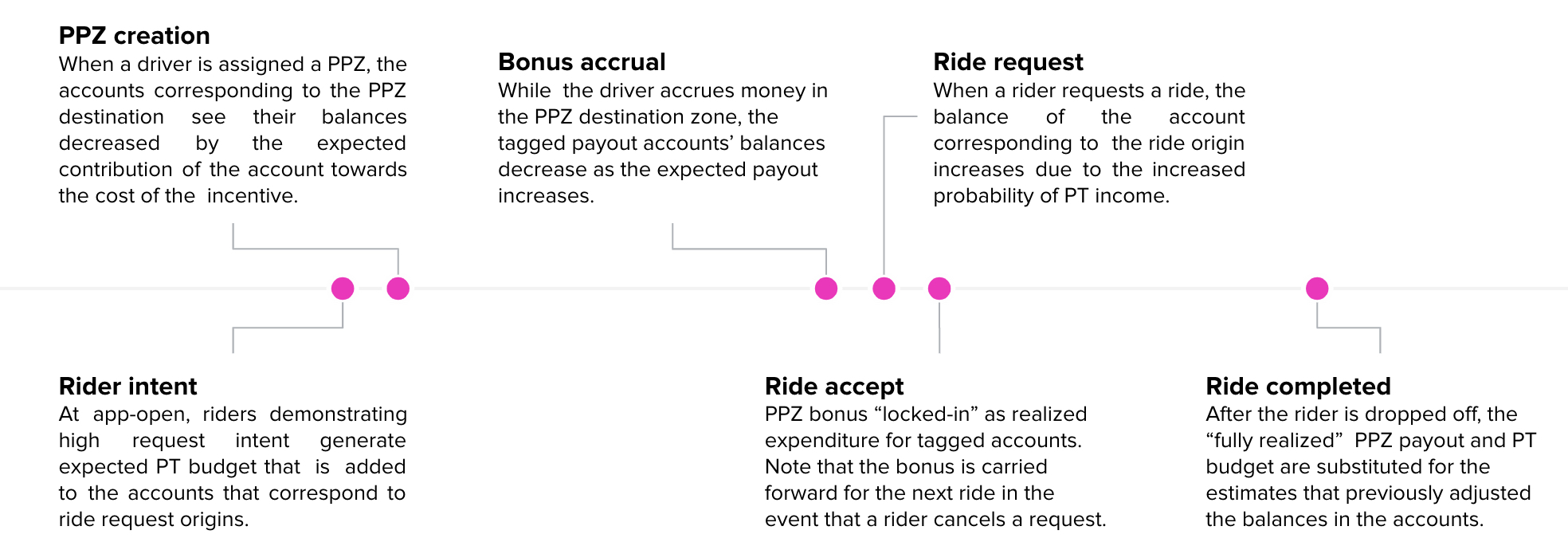}
    \caption{A sample timeline illustrates how riders generate escrow account balances that allow PPZs to be generated and served to encourage drivers to reposition. Note that these sample events illustrate how the escrow accounts are updated for expected incoming budget and payouts and eventually ``realized'' as finalized account balances.}
    \label{Figure4}
\end{figure}

PT income is attributed to the local account corresponding to the location at which the ride request originates. Based on the repositioning optimization, PPZ expenditure is attributed to a set of local accounts corresponding to the PPZ destination. A subtlety is that PT income is attributed to a single account whereas PPZ expenditure can possibly be attributed to multiple accounts near the PPZ destination. The reason for this is that the driver who reaches a PPZ destination may subsequently be dispatched to pick up riders in locations close to the destination as well as those in the destination location. Thus, pooling multiple account balances to motivate the driver to reposition is appropriate. Although the full set of events causing updates to the account balances is too numerous to list, \Cref{Figure4} gives an intuitive illustration of how riders and drivers typically interact with the platform and consequently induce escrow account balance changes.

As we explain above, we want to avoid accumulating money in the accounts (as opposed to spending it in real time); to avoid this, the PPZ bonus amounts offered are set to equal approximately the ratio of the available balance in the accounts to the number of drivers who are expected to reposition to the locations tied to the local escrow accounts, given the PPZ incentive. They thus form a set of ``account clearance targets'' that functions as a budgetary reference for an amount to spend at each time step of our PPZ allocation and bonus computation. Its purpose is to clear our accounts and avoid accumulating money. Such target spend signals from the escrow mechanism provide a safe way for PPZs to spend incoming PT money quickly without overspending. In addition, they serve as principled, location-based accounting signals for PPZ to incentivize drivers to reposition.

\subsection{Problem Decomposition and Certainty Equivalent Approximation}

The problem described in the \emph{Stochastic Model} section is a Markov decision process (MDP) that can theoretically be solved optimally; however, such an approach is impractical due to the curse of dimensionality. To avoid the high dimensionality, we first decouple the problems of (1) determining which drivers to incentivize to reposition, and (2) determining the size of the bonus to pay each driver for having repositioned. Before we provide a high-level description of these two problems, we next describe our use of a certainty equivalent approximation to the repositioning problem.

The certainty equivalent approximation replaces the assumption that drivers act stochastically; that is, the~$n$ drivers receiving a PPZ decide to relocate from one location to another independently based on a coin toss with bias p, with the assumption that exactly the expected number of drivers will relocate (i.e., $np$). Under this approximation, we can replace the high-dimensional MDP by an optimization problem that can be convexified. Applying the concepts described in \emph{Practical Implementation Details} in \emph{Appendix A: Optimization Approach}), the convex optimization problem can be solved using commercial solvers within seconds to a high degree of accuracy. As is the case in similar models (\cite{braverman2016empty}, \cite{banerjee2016pricing}, \cite{ozkan2016dynamic}), one can show that in commonly studied large market settings the solution to the convexified certainty equivalent problem approximates the intractable optimal solution to the MDP. This is based on a standard concentration argument wherein we construct a fluid upper bound, and show that as the market grows large, the relative gap to that upper bound becomes small.

\subsection{Sequential Algorithm}

\begin{figure}[t]
    \centering
    \includegraphics[width=\linewidth]{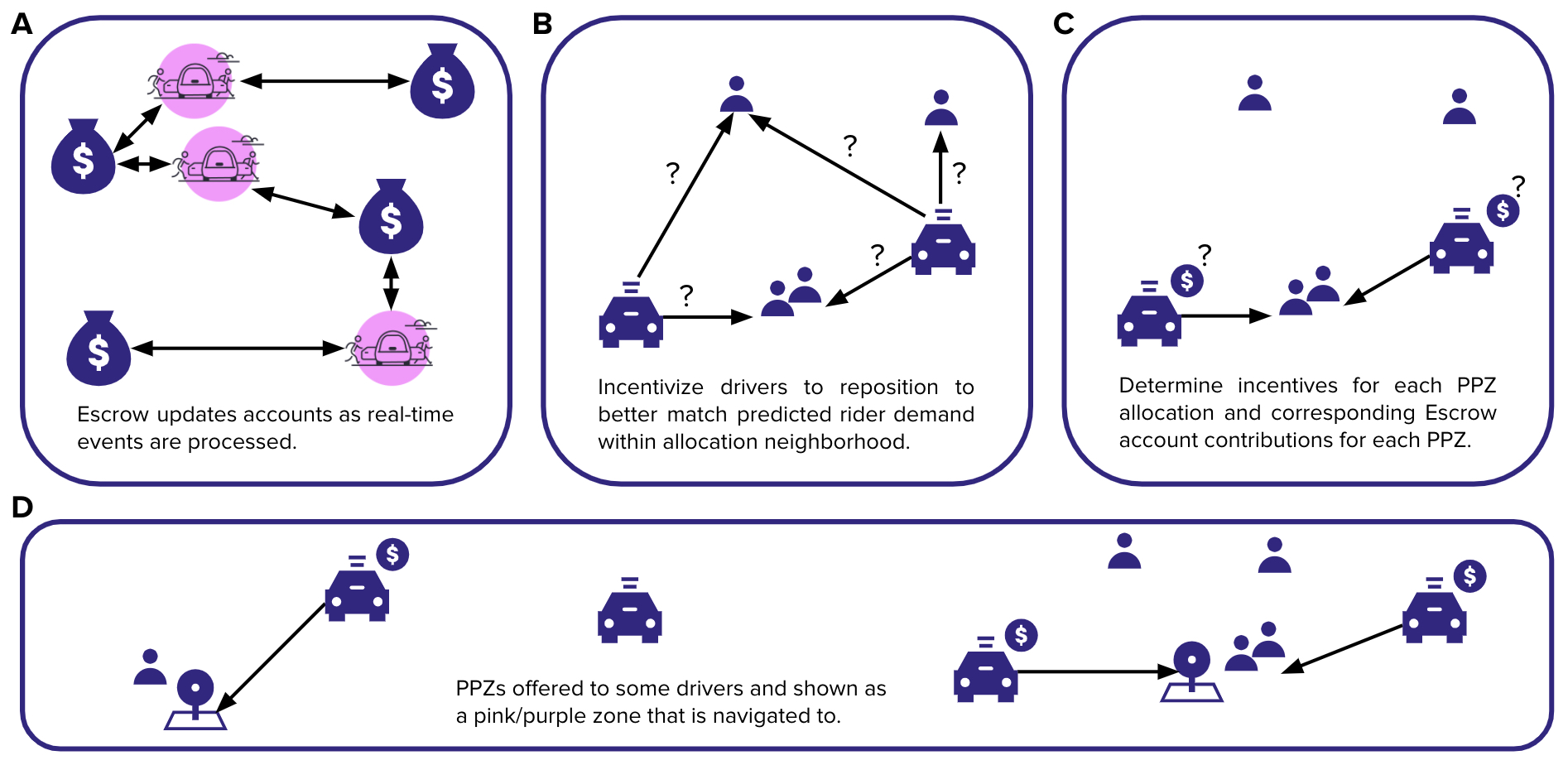}
    \caption{The graphic shows a systemic view of the driver positioning and incentive computation subroutines, which are executed for a city to improve the spatial distribution of drivers and match rider demand.}
    \label{Figure5}
\end{figure}

\Cref{Figure5} illustrates how the escrow mechanism’s account balances are updated (A) and entered into the pair of subproblems (B and C), which are then solved to obtain the PPZ allocations sent to drivers at each time step (D). First, in (A), PT income and PPZ payouts are distributed to and contributed from various escrow accounts depending on driver and rider events and the locations at which they occur. The escrow account updates occur in real time and the account balances are entered as parameters to the PPZ optimization subroutines. Then, in (B), the algorithm optimizes for the relocation of drivers to match the locations at which future rider requests are expected to originate while ignoring the exact payouts needed to incentivize drivers to reposition. In (C), the incentive computation subroutine produces actual bonus payout values for the PPZ to be given to each driver. Finally, in (D), PPZs are created to incentivize a subset of the idle drivers to reposition and better serve predicted rider requests. To maintain some level of parity between PT income and driver bonuses based on where the ride requests originate, we enforce an equal-split payout among all drivers guided to the same PPZ destination.

\subsection{Driver Positioning}

Our driver positioning subproblem is to maximize the expected revenue-weighted rider request ``conversions'' in the entire city. The key decision variable is the set of PPZ allocation fractions $A_{ij} \in \left[0,1\right]$ from some location~$i$ with open drivers to another location~$j$. These allocation fractions tell us how many drivers we want to incentivize in any one location to reposition to another location (in a single decision period). More explicitly, for some driver in location~$i$, we choose some destination out of all locations available~$j$ weighted by~$A_{ij}$ (including the ``null location'' in which we do not serve any PPZ).

The PPZ allocation fractions are set based on constraints on the service level, the dynamics of how drivers reposition vis-a-vis PPZ incentive allocations, and budgetary restrictions arising from the available escrow account balances. The service level constraint ensures that demand (subject to pricing) does not exceed supply in the location of the spatial neighborhood to which drivers may be dispatched. That is, in each location’s ``dispatch neighborhood,'' the market is \emph{balanced} by simultaneously inducing PT and accounting for the drivers incentivized to reposition. A notable aspect of the budgetary constraints is that although local account balances are aggregated and tagged to discrete locations, they are accessible to PPZ offers that guide drivers to any destination location within a spatial ``contribution radius'' around the local account's location tag. This design choice for the escrow mechanism is motivated by the fact that upon reaching a specific PPZ destination, a driver can subsequently be dispatched to not only the destination location’s riders but also to riders in nearby locations. The exact details are provided in \emph{Appendix A: Optimization Approach}. Here, we simply highlight that this property allows us to make flexible allocation decisions; for example, we can serve forecast demand located near but not at the locations of existing demand.

\subsection{Incentive Computation}

Given the allocations from the driver positioning subroutine, the incentive computation subproblem attempts to set the monetary payout so that every driver sent to the same location is offered the same expected bonus. The rationale is as follows. The balances at each location represent the sum of the PT markups that each rider will pay for trips of different lengths. Although we cannot know which ride request will be dispatched to a driver at some location ahead of time, one fair approach is to average the expected PT income over all possible trips to which that driver can be dispatched from that location. This results in an even split of the expected PT income among drivers. One decision variable is thus the set of per-PPZ bonus values unique to each destination location. Another decision variable is the set of contribution fractions indicating the portion of the available balance in each escrow account to be set aside for every PPZ offer ending at some destination location. These decisions are constrained by simple lower and upper bounds on the range of permissible bonus values. An important constraint is that the expected expenditure resulting from drivers who respond to the PPZ instructions should not exceed the available budget.

\section{Numerical Experiments}

We carry out numerical simulation using previous data in back-tests to exercise our algorithm in a safe nonproduction environment. The back-tests also enable us to plan for required computation resources and understand how the model handles specific scenarios, for example, how it handles peak and nonpeak commute periods or cities that clearly divide their downtown and suburban areas and those that do not.

\subsection{Data}

We work with three months of data from cities (anonymized for confidentiality) that are representative of the diversity of the topographies of the cities in which we operate. We collect data for problem parameters from production logs. As a proxy for the account balances for which we do not have historical data, we assume that the system will always clear the account balances at each decision period. As a result, the balances available at each decision period come from the historical PT transactions from the previous period.

\subsection{Benchmarking Simulation}

\begin{figure}[t]
    \centering
    \includegraphics[width=0.7\linewidth]{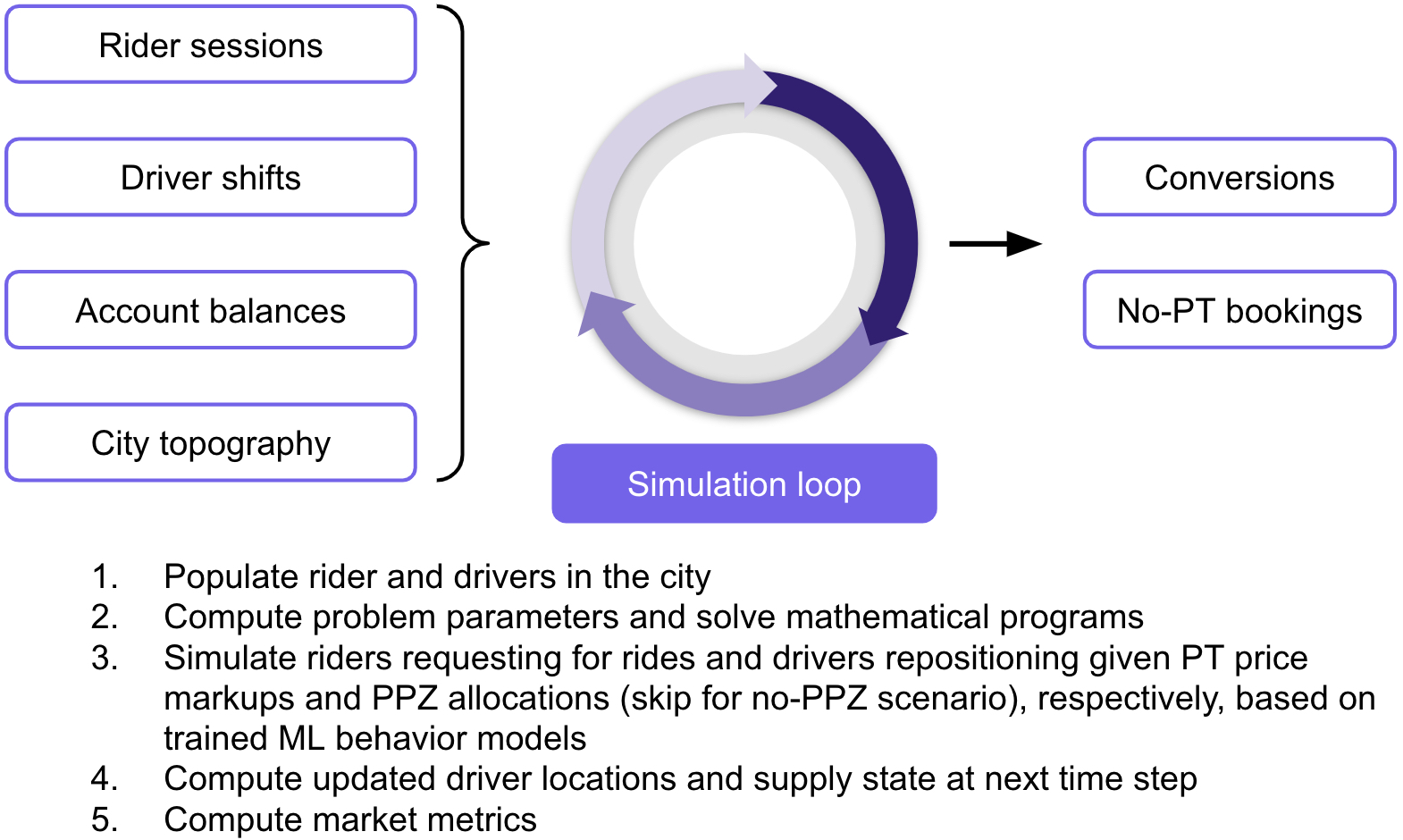}
    \caption{
    Initial market conditions are sampled and propagated for one time step to obtain performance metrics such as rider conversions and no-PT bookings.
    }
    \label{Figure6}
\end{figure}

We consider a driver positioning and incentive budgeting strategy compared against a null allocation benchmark. The null allocation benchmark represents the legacy market conditions with the driver PT product without PPZ allocations. Rather than a comprehensive rollout of the PPZ model over a long time horizon, we consider only the results of our single-period model, as shown in \Cref{Figure6}. Thus, our simulation cannot account for the cumulative impact of PPZs incentivizing the drivers’ repositioning in previous time steps and will be biased toward returning higher-impact estimates.

We conduct back-test simulations for four cities, using two strategies that optimize for different objectives, namely, the rider request conversions and no-PT bookings. For each simulation, we give a relative version of the incremental objective value or return. We do not directly use the objective value from the optimization. Instead, we take the allocation results and enter them into a rudimentary simulator to obtain the performance metrics. The simulator employs a model fitted from historical data (e.g., idle behavior and dispatch likelihood) to determine the response of the drivers to the allocation guidance and the simulated return.

\Cref{Table1} shows that the performance of the algorithm and hence product can vary significantly as the city changes. Empirically, cities that are less homogeneous in population density across the region tend to enjoy greater metric improvements (e.g., cities A and C in our simulations). Some examples are cities with sharp concentrations of demand at residential and office buildings at various times of the day. This often means that there are greater opportunities in incentivizing drivers to move between locations to improve the market balance. This observation is consistent with our goal of rewarding drivers that reposition themselves with PPZs.

\begin{table}[t]
    \centering
    \caption{The table illustrates simulation results with different objectives and cities and the relative incremental return values.}
    \label{Table1}
    \begin{tabular}{lc|cccc}
        \toprule
            Objective & City & \multicolumn{2}{c}{Conversion} & \multicolumn{2}{c}{No-PT bookings} \\
                      &      & Mean & Median & Mean & Median \\
        \midrule
            Conversion & A & 2.350\% & 1.504\% & 1.151\% & 0.622\% \\
                       & B & 0.758\% & 0.092\% & 0.222\% & 0.022\% \\
                       & C & 4.254\% & 2.009\% & 0.364\% & 0.075\% \\
                       & D & 0.905\% & 0.114\% & 0.426\% & 0.021\% \\
            No-PT      & A & 0.913\% & 0.499\% & 1.902\% & 1.179\% \\
            bookings   & B & 0.205\% & 0.021\% & 0.705\% & 0.097\% \\
                       & C & 0.426\% & 0.064\% & 3.601\% & 1.450\% \\
                       & D & 0.346\% & 0.002\% & 0.466\% & 0.003\% \\
        \bottomrule
    \end{tabular}
\end{table}

\begin{figure}[t]
    \centering
    \begin{subfigure}{.5\linewidth}
        \centering
        \includegraphics[width=\linewidth]{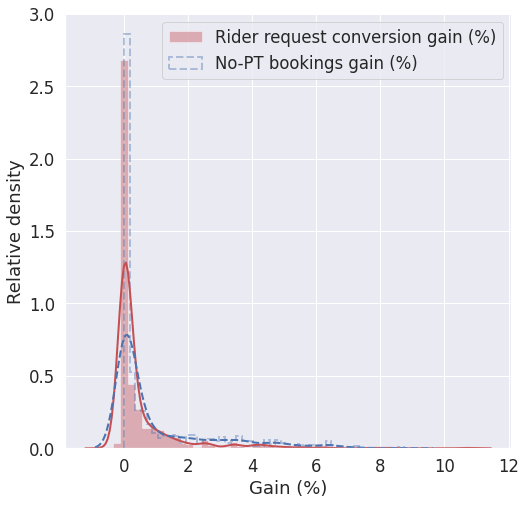}
    \end{subfigure}%
    \begin{subfigure}{.5\linewidth}
        \centering
        \includegraphics[width=\linewidth]{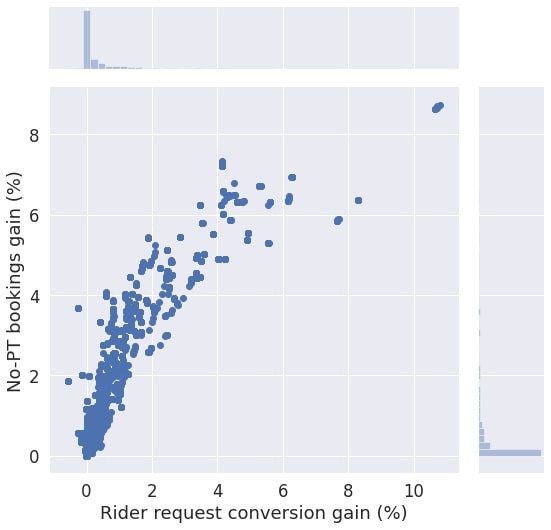}
    \end{subfigure}
    \caption{In the distribution (left) and joint (right) plots of the incremental gain in no-PT bookings and rider request conversions, the plots are normalized such that the area under the curves totals 1. The joint plot of the incremental gains (right subfigure) show that the gain in no-PT bookings (the objective) is about twice that of the rider conversion gain and that optimizing for either metric still improves the other.}
    \label{Figure7}
\end{figure}

\begin{figure}[t]
    \centering
    \includegraphics[width=0.8\linewidth]{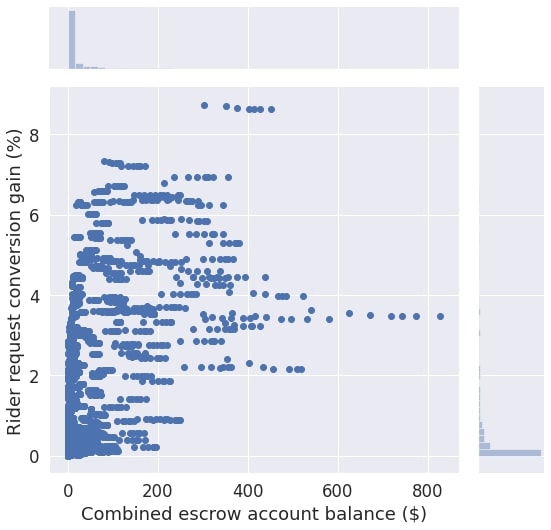}
    \caption{
    The right skew in the distribution of gains can be explained because randomly sampling timestamps uniformly overwhelmingly returns timestamps with no-PT rides and therefore no escrow budget.
    }
    \label{Figure8}
\end{figure}

To illustrate the above, \Cref{Figure7} plots the relative incremental gain results for City A based on the no-PT bookings maximizing algorithm. The distribution plot (left subfigure) shows a right skew in the incremental gains. This skew is due to our uniformly random sampling of periods causing our distributions to be dominated by lull periods, as opposed to demand peaks with greater repositioning opportunities (see \Cref{Figure8}). We remark that the significant performance gains on the y-axis when the budget is extremely small can be explained by scenarios in which the driver count is low, such that only one PPZ with, for example, \$3 can yield relative gains in rider request conversion of a few percentage points.

\section{Live Experiments}

We ran a series of experiments with the twin goals of estimating the marketplace outcomes of our product and determining whether to launch it to all Lyft markets. In particular, we were interested in estimating the effect of our incentive-based supply positioning system on conversion, that is, the probability that if a potential rider opens the Lyft app, that opening becomes a Lyft ride. In the context of our marketplace, naive estimators suffer from network interaction or interference bias, which can result in estimates that differ significantly from actual effects, sometimes by an order of magnitude \cite{Cha16}. We briefly introduce the techniques we used to circumvent this problem before presenting the results of our experiment.

\subsection{Causal Inference Framework}

The most widely used causal inference technique to circumvent the problem of statistical interference in ridesharing platforms and marketplaces, including Lyft, is time-split or switchback tests \cite{XLG+18}. Treatment is randomly assigned by time interval (rather than user) so that all the users in the market in a given time interval belong to only one variant and the effect of interference is minimized. Unfortunately, we could not use switchbacks because (1) our product involved a major change to the user interface and switching back and forth between two substantially different experiences for the same user was not feasible, and (2) we wanted to measure the long-term effects of continuously exposing users to our product. For these reasons, we based our inferences on a driver-split experiment in conjunction with a model that we built and validated to correct for most of the interference bias. 

We wanted to estimate how a change in the distribution of supply would affect the aggregate number of rides that can be dispatched on the platform. At the micro-scale, this is a function of the effect of local supply on the probability that a rider opening the Lyft app becomes an actual ride. Our model obtains this effect from experimental data by (1) heuristically estimating the counter-factual supply that would occur in a scenario with 100\% treatment drivers or with 100\% control drivers, and (2) using a machine learning model to map the counter-factual supply to incremental conversion (or, indeed, any market-level metric such as revenue, bookings, or ETAs). We then aggregate these micro-scale estimates to obtain the macro-level effect across the platform.

Suppose we run a 50-50 driver-split test in which we assign half the drivers the old PT experience and half the drivers the new PPZ experience. For most drivers on our platform, the local supply will include both treatment and control drivers. We compute the counter-factual supply by rescaling the number of drivers in a group as if there were only drivers from that group. For example, if a rider had 5 nearby treatment drivers and 3 nearby control drivers, we compute 100\% control supply as 3 × 2 = 6 and 100\% treatment supply as 5 × 2 = 10. We make an additional adjustment that takes into account the way rides are matched and the riders’ request elasticity with respect to supply.

We then combine our estimates of counter-factual supply with a conversion model. This is a machine learning model that was trained offline and estimates the probability that a rider requests a ride, given the observed local supply, observed local demand, plus other context features such as time and location. It is a flexible model that we can use to estimate the direct impact of the experiment by comparing the factual scenario to the 100\% control counter-factual and the potential impact of rolling out the product by comparing the 100\% treatment counter-factual to the 100\% control counter-factual. In both cases, the incremental conversion estimates are interpreted as causal, because they hold constant all the variables, other than local supply, that affect rider conversion.

As we note above, the goal of this approach is to remove most of the interference bias and produce estimates that are close to a bias-free, but in our case, infeasible experimental design. Some amount of bias is unavoidable, unfortunately. For example, when testing PPZ and the legacy system in a 50-50 driver-split test, the drivers in the legacy system observe prices that are set for all riders; that is, they depend partly on the positioning of the treatment drivers. To the extent that drivers chase the surge, which expert drivers recommend against as we mention above, this causes some interference bias in our results. Fortunately, our validation experiments showed that this residual bias is small compared to the interference bias for which our framework successfully corrects.

We validated our framework with a mixed time-split and driver-split design using a driver positioning incentive as treatment, which has similar marketplace effects to our PPZs but does not involve a drastic user-interface (UI) change and thus permits time-splits. The goal of the validation was not to create inference about the effectiveness of PPZs but simply to verify that, in a similar setting, our causal inference framework produces estimates that are close to the interference-free time-split. The experiment involved randomizing treatment daily and by users: in off days no one was assigned treatment, and in on days, 50\% of the users were randomly assigned treatment. The results were impressive: the time-split estimated an incremental bookings treatment effect of +1.6\% from our intervention, whereas the driver-split naive estimate (taking the difference between the treatment and control variants) was much higher at +39.7\%. (The naive estimate is highly inaccurate because of a cannibalization effect: consider a high-demand area with a sufficient number of drivers to serve all requests. PPZs offered to encourage additional drivers to relocate to this area do not increase the number of riders served; however, they increase the percentage of riders served by treatment drivers, and thus cannibalize the would-be riders for the control drivers.) In contrast, our causal framework gave an estimate of +1.2\%, which is close and statistically indistinguishable from the time-split estimate. This gave us confidence in the effectiveness of our causal inference framework.

\subsection{Live Experiments: Results}

\paragraph{Bookings.} We implemented and tested the version of the PPZ model that optimizes for no-PT bookings in selected cities. Experimental results show that the performance improvement brought by the PPZ algorithm is consistent in all cities, with gains in the bookings averaging at slightly higher than 0.5

\paragraph{Other marketplace improvements.} We observed a reduction in ride pick-up times and, correspondingly, drivers spending more time in their cars with riders. We also observed improvements on a number of driver engagement metrics. (1) We saw a small but significant increase in active drivers for the drivers with PPZ versus the ones with the legacy system during the test period (+0.82\%), (2) drivers drove more hours with PPZ in both nonpeak and peak hours (+0.53\%), with the biggest lift concentrated during peak hours (+1.43\%), and (3) ETAs (-1.1\%) and driver cancel rates (-12.5\%) were down significantly (thus validating PPZ’s effectiveness on addressing the problem of drivers cherry-picking rides).

\paragraph{Driver earnings and driver sentiment.} Driver earnings were constant in aggregate. This is expected because we are only shifting the budget allocation and increased earnings would lead to increased supply, thus lowering driver utilization, and thereby decreasing earnings, again, in equilibrium \cite{hall2021labor}. However, the distribution of earnings changed in a way that reduces the inequality induced by differences in experience levels (i.e., less experienced drivers are more productive with PPZ). In particular, median hourly driver earnings increased by \$0.20 with PPZ, and by up to \$0.50 during some weeks. Surveys sent to drivers who have experienced the product also show a preference for the PPZ product over PT across different cities. Some responses to the new system include the following quotes submitted by anonymous drivers in these surveys:

\emph{``Before, I used to drive looking for my next pick up. Now I can stay in the area waiting for my next pick up and the good thing is I’m waiting and earning money, thanks!''}

\emph{``LA is so big, nice to not waste time!''}

\emph{``It’s a great incentive and makes me want to go to areas to earn more on my rides.''}

In addition, we also received positive reviews from operations teams working in our major markets.

\emph{``PPZs increased utilization and hours. I think this was communicated well, intuitive for drivers. Performing well thus far—all driver metrics look positive, and have not seen significant adverse sentiment in-market.''}

Given these promising results, the proposed algorithm and product have been successfully deployed on Lyft’s ridesharing platform across all 320 cities. See \emph{Appendix C: Video of PPZs Served in Production} for an illustration of the product being served ``live'' in the San Francisco Bay Area.

\section{Broader Impact}

The introduction of mathematical optimization and, more generally, operations research techniques in this application has paved the way for more sophisticated analysis and decision making in incentive budgeting and generation. That Lyft already uses operations research techniques in many of its core applications, including order dispatch and trip pricing, is no surprise. That said, incentive budgeting and generation for drivers has previously been accomplished using heuristics. PPZ represents the first time that dynamic decision making in real time has been automated with mathematically principled algorithms.

Beyond the PPZ product, the escrow mechanism we built is being developed to provide complementary real-time incentives that can access the same hundreds of millions of dollars of yearly budget. In particular, the escrow mechanism allows us to funnel part of the budget to other incentives that can act in tandem with PPZ to achieve even better market results. For example, as we can see from our live-test results, PPZ does not increase the participation rates of drivers in the short term; that is, it is not designed to encourage offline drivers to come online to drive on the platform, nor does it significantly impact them to do so. One direction that we are actively exploring is adapting weekly, manually managed incentives, which already exist for drivers (e.g., ride streaks), to encourage more drivers to come onto the platform when Lyft experiences an unexpected driver supply shortage.

We are currently working on some of these applications, stretching our system to account for different types of supply and testing new products that work in concert with PPZs. These new applications will help engage online drivers and signal earnings opportunities to offline drivers, with the broader goal of increasing market balance and efficiency on the platform.

\section*{Acknowledgments}

We thank David Shmoys, Siddarth Patil, and Chris Sholley for their feedback on this paper and prior drafts. PPZ is still a work-in-progress, and the work would not have been possible without the support of many people and various teams at Lyft. The core group of researchers who helped develop the PPZ algorithm, the escrow mechanism, and its dependencies include Ido Bright, Cameron Bruggeman, Carolyn Cotterman, Benedict Kuester, Michael Rotkowitz, Lei Tang, and Michael Yoshizawa. The product has benefitted greatly from the feedback of our colleagues at Lyft and Lyft’s users. In particular, we thank the following people for their contributions to PPZ: Ben Dear, Eduardo Apolinario, Matt Green, Dan Barragan, Praveen Athmanathan, Richard Zhao, Gaurav Gupta, Seth Melnick, Bryan Jung, David Linder, Efferman Ezell, Vijay Narasiman, Eli Schachar, Jia Yan, Ramon Iglesias, Varun Krishnan, Charlene Zhou, Akshay Balwally, San Tan, Udi Milo, Jose Abelenda, Jeremy Karp, Derek Salama, Adriel Frederick, John Fremlin, and Garrett J. van Ryzin.

\newpage
\bibliography{template}
\newpage

\appendix
\section{Optimization Approach}

The PPZ algorithm has two main subroutines: the driver positioning and incentive computation subproblems.
We first define the notation and provide details of our model, which is independent of the methods for the allocation of PPZs or the evaluation of the platform efficiency.
We then describe the optimization problems that we solve to generate PPZ allocations and the dollar bonus values offered for each PPZ.
We refer the reader to the main text for a high-level overview of the system.
We highlight that, to focus on the core challenges of the PPZ problem, we leave out an important component of the PPZ system.
In particular, we assume that we have access to generic problem parameters that are used across many different teams at Lyft.
This includes, among others, forecasts and estimates of supply, demand, and other important market conditions.
Below, all defined quantities are given as forecasts or estimates of the market conditions unless otherwise stated.

\subsection{The Model}

\subsubsection{Order dispatch.}

Ridesharing platforms typically compute dispatches across a city or region, and Lyft is no exception.
A balanced market has an adequate supply of drivers for the riders requesting ride dispatches (i.e., demand) at the geographic neighborhood level around any location.
We call this the dispatch neighborhood.
Consider a contiguous spatial region with $n$ nonoverlapping, discrete spatial clusters or locations.
We denote the ``local'' demand and idle supply as $\left(d^\text{req}\right)_{1:n}$ and $\left(s\right)_{1:n}$, respectively, for all locations $i = 1, \ldots, n$.
The ``neighborhood'' demand and idle supply around all locations $i = 1, \ldots, n$ are denoted as $\left(d_{N}^\text{req}\right)_{1:n}$ and $\left(s_{N}\right)_{1:n}$, respectively.
To obtain the neighborhood quantities from the local ones, we introduce the dispatch neighborhood matrix $M \in \left\{0, 1\right\}^{n \times n}$, where each entry $M_{ij}$ indicates whether locations $i$ and $j$ are within the same dispatch neighborhood.
Using this matrix, we select and add up the demand and supply within the neighborhood of each location (\ie, $d^\text{req}_{N} = M d^\text{req}$ and $s_{N} = M s$).
In practice, the neighborhood matrix depends on region-specific dispatch parameters and prevailing traffic conditions.
It is also very sparse.
We also, by ride prices and incentives, try to maintain a level of available drivers $\left(r\right)_{1:n}$ in the dispatch neighborhood of each location $i = 1, \ldots, n$ to avoid a \emph{wild goose chase} scenario in future periods.
As \cite{CKW17} explain, a platform depleted of idle drivers is often required to match drivers with riders who are far from where the driver is positioned.
These chases occupy drivers and reduce the rate of rides served and earnings, which exacerbate the problem.
Through PT and PPZ, we simultaneously modulate demand and incentivize drivers to position themselves to maintain a healthy level of available drivers in the marketplace.
We express the market-balancing condition as
\begin{equation}
    d^\text{req}_{N} = M d^\text{req} \preceq M s - r = s_{N} - r\text{.}
\label{eqn-model-balance}
\end{equation}

\subsubsection{Price modifiers and elasticity.}

The PT service determines the price modifiers $\left(x\right)_{1:n}$ to be applied to rides at each location $i = 1, \ldots, n$.
We model the aggregate rider response to the price modification at each location using a conversion function, which is given by $y: \reals^n_+ \rightarrow \left[0, 1\right]^n$.

It gives the probability that a rider (unit demand) will request a ride at some location, given some PT value.
Denoting the number of riders on the application determining whether to request a ride as $d \in \reals_+^n$, the expected PT-modulated demand is $d^\text{req} = d \circ y\left( x \right)$, where $\circ$ denotes the Hadamard (elementwise) multiplication of vectors.
Based on observed data, conversion exponentially decays with increasing prices.

\subsubsection{Driver allocation.}

The PPZ system determines the driver allocations $A \in \reals^{n \times n}$, where $A_{ij}$ is the fraction of drivers in location $i$ to whom we serve incentives to encourage them to move to location $j$.
Naturally, the driver allocation fractions originating from each location can sum at most to unity.
We limit the repositioning incentives of drivers to within a geographic neighborhood of their original location to avoid placing the incentives too far away from the driver or at a location that cannot be reached (\eg, construction zones).
We encode undesirable allocations by restricting the appropriate elements to zero in $A^\text{max} \in \left\{0, 1\right\}^{n \times n}$.
Our driver allocation constraint is
\begin{equation}
    A\ones \preceq \ones, \quad 0 \leq A \leq A^\text{max}\text{.}
\label{eqn-model-allocsimplex}
\end{equation}

\subsubsection{Supply dynamics.}

We model the evolution of supply dynamics through the probability of a driver transitioning to a specific location $k$ at the next period $P_{k, ij}$, given a PPZ to go from the driver's current location $i$ to the location $j$.
Drivers without PPZs stay open with probability $P_0$.
Suppose~$\left(s_0\right)_{1:n}$ is the pre-PPZ allocation driver count at locations $i=1,\ldots,n$.
Then, the expected supply evolves as
\begin{equation}
s = \begin{bmatrix} \ones^T \left( P_{1} \circ A \right ) s_0 \\ \vdots \\ \ones^T \left( P_{n} \circ A \right ) s_0 \end{bmatrix} + P_0 \left( I - \diag\left( A \ones \right) \right ) s_0 \text{.}
\label{eqn-model-supply-dynamics}
\end{equation}
To unpack this equation, recognize that $P_k \circ A$ gives the probability that each allocation will result in a transition to location $k$.
Therefore, $\left( P_k \circ A \right) s_0$ gives the vector of drivers that, in expectation because of how we sample PPZ allocations for each driver, end up in location $k$ from each origin location $1, \ldots, n$.
Summing the elements of this vector produces the expected number of drivers that end up in location $k$.
Next, recognize that $A \ones$ gives the total fraction of allocated drivers from each origin such that $I - \diag\left(A \ones\right)$ gives the unallocated fraction of drivers at each location.
It now follows that $\left(I - \diag\left(A \ones\right)\right) s_0$ is the vector of driver counts that were unallocated at each location; multiplying it by $P_0$ gives the probability that they stay on the platform.
Defining a response probability matrix $P_{c} \in \reals_+^{n\times n}$ with elements $P_{c, ij}$ denoting the probability that a PPZ with origin $i$ and destination $j$ will be satisfied, the expected number of drivers that satisfy and earn their PPZs can be written as $\left(P_c \circ A\right) s_0$.
Notice that $P_c$ is related to $P_k$ in that $P_{c, ij}$ is simply the subelement $P_{j, ij}$.

\subsubsection{Escrow budgeting.}

The escrow mechanism uses real-time PT and PPZ financial line items at ride accept to track the available budget to spend $\left(e\right)_{1:n}$ at each location $i=1,\ldots,n$.
For any PPZ, the escrow mechanism requires a set of contributing accounts and contribution fractions that funds it.
This set of contributions can be encoded by $C\in\reals^{n\times n}$, where $C_{ij}$ is the fraction of budget contribution from location $i$'s escrow account to any PPZ with the destination location $j$.
Similar to the allocation matrix constraint, we restrict invalid contributions to some location from a local account that is too far away from zero in $C^\text{max} \in \left\{0, 1\right\}^{n\times n}$.
The contribution constraints
\begin{equation}
C \ones \preceq \ones, \quad 0 \leq C \leq C^\text{max}
\label{eqn-model-contribution-simplex}
\end{equation}
encode the fact that the (valid) budget contribution fractions must sum at most to unity.
We require that the total amount of money allocated for each PPZ destination location cannot exceed the total available from the local balances $C e$, even if we offered the minimum allowable bonus amount $b^\text{min}$ to every responding driver $\left(P_c \circ A\right) s_0$.
The contributions are thus governed by
\begin{equation}
b^\text{min}\left(P_c \circ A\right) s_0 \preceq C e\text{.}
\label{eqn-model-budget-contribution-bound}
\end{equation}

For any valid allocation $\tilde{A}$, the final bonus values shown to drivers are governed by
\begin{equation}
\diag\left(\left(P_c \circ \tilde{A}\right) s_0\right) b \preceq C e, \quad b^\text{min} \preceq b \preceq b^\text{max}
\label{eqn-model-budget-bonus-bound}
\end{equation}
such that the total payout for all complied allocations in each destination location is not more than the budget available in the escrow account balances.
Further, all bonus values are subject to lower and upper bounds $b^\text{min}$ and $b^\text{max}$, respectively.

\subsection{Driver Positioning}

The driver positioning objective is to maximize the aspects of profit that PPZs can directly impact.
Specifically, we maximize the expected no-PT bookings $f^T \left( d \circ y\left(x\right) \right )$, where $\left(f\right)_{1:n}$ are the time-and-distance fares we expect to collect at locations $i=1,\ldots,n$.
Our key decision variable is the allocation $A$ and the free variables are the PT values $x$, the open supply that does not already have a PPZ $s$ and thus can be offered an incentive to reposition, and the escrow contribution $C$.
Note that $\bar{s}$ is a parameter that gives the count of drivers in each location that already have a PPZ and thus cannot receive an incentive to be repositioned.
The positioning problem is to maximize the expected bookings subject to the market-balancing condition [Equation~\eqref{eqn-model-balance}], supply dynamics [Equation~\eqref{eqn-model-supply-dynamics}], driver allocation constraints [Constraints~\eqref{eqn-model-allocsimplex}], and escrow budgeting constraints [Constraints~\eqref{eqn-model-contribution-simplex} and \eqref{eqn-model-budget-contribution-bound}].
Notice that, from our certainty equivalent approximation, the optimization assumes that the supply and demand dynamics evolve as in expectation.
Our problem is
\begin{equation}
\begin{array}{ll}
\mbox{maximize} & f^T \left( d \circ y\left(x\right) \right ) \\
\mbox{subject to} & M \left(d \circ y\left(x\right)\right) \preceq M \left(s + \bar{s} \right ) - r \\
& b^\text{min} \left( P_c \circ A \right ) s_0 \preceq C^T e \\
& A \ones \preceq \ones, \quad 0 \leq A \leq A^\text{max} \\
& C \ones \preceq \ones, \quad 0 \leq C \leq C^\text{max} \\
& s = \begin{bmatrix} \ones^T \left( P_{1} \circ A \right ) s_0 \\ \vdots \\ \ones^T \left( P_{n} \circ A \right ) s_0 \end{bmatrix} + P_0 \left( I - \diag\left( A \ones \right) \right ) s_0 \text{,}
\end{array}
\label{equation-driver-positioning-problem}
\end{equation}
which yields the optimal set of PPZ allocation fractions $A^\star$.
In practice, we add an $\ell_1$ regularization term on the allocation to induce sparsity such that the set of possible PPZ allocations is small for each origin location.
We also apply an $\ell_2$ penalty term onto the vector of differences in the price variables in adjacent locations to induce spatial smoothness.

\subsection{Incentive Computation}

For each location $i$, the escrow mechanism provides a bonus target $b_i^\text{tgt}$ that an open driver should ideally receive by averaging out the available account balances surrounding $i$.
Recall that this is the chosen fair approach of splitting the marked-up income by averaging it out over all possible trips for which drivers can be dispatched from each destination.
Given the optimal PPZ allocation $A^\star$, the incentive computation subproblem attempts to find escrow account contributions $C$ to match the bonus values $b$ with the ideal bonus target $b^\text{tgt}$ such that $\left\| b - b^\text{tgt} \right\|_2^2$ is minimized.
The bonus targets are set such that the available budget is fully spent.
The solution is subject to the bonus value constraints [Constraints~\eqref{eqn-model-budget-bonus-bound}] and the contribution simplex [Constraints~\eqref{eqn-model-contribution-simplex}].
Note that simply setting the final bonus payout values to the ideal bonus target will not necessarily yield a feasible solution under the budget constraints.
In particular, the minimum and maximum bonus values sometimes restrict us from using the ideal budget target as solutions.
The problem is
\begin{equation}
\begin{array}{ll}
\mbox{minimize} & \left\| b - b^\text{tgt} \right\|_2^2 \\
\mbox{subject to} & \diag\left( \left(P_c \circ A^\star \right ) s \right ) b \preceq C^T e \\
& b^\text{min} \preceq b \preceq b^\text{max} \\
& C \ones \preceq \ones, \quad 0 \leq C \leq C^\text{max} \text{,}
\end{array}
\label{equation-incentive-computation-problem}
\end{equation}
which yields the optimal bonus payouts $b^\star$ and escrow contribution $C^\star$.
In practice, we apply an $\ell_1$ regularizer on the contributions $C$ such that the set of accounts contributing to any PPZ is not excessively large.

\subsection{Practical Implementation Details}

We now describe some necessary steps that allow the optimization to run efficiently at Lyft’s scale.

\subsubsection{Convexification.}

As presented, the PPZ optimization problem is nonconvex, which greatly complicates its solution and, for all practical purposes, makes its computation potentially unfeasibly long.
The nonconvexity is the result of the conversion function that appears in the adjusted market-balancing condition [Equation~\eqref{eqn-model-balance}] and the objective function.

Fortunately, a change-of-variable convexifies the problem.
Since $f$ is strictly monotone, we can let $y = y \left( x \right)$ (in a slight abuse of notation) and recover $x$ by inverting the conversion function.
Observe that the new variable has domain $y \in \left[ 0, 1 \right]$ and that each element $y_i$ is strictly decreasing in $x_i$ for $i = 1, \ldots, n$.
With this change-of-variable, Equation~\eqref{eqn-model-balance} and the objective function become affine.
Our trick here resembles the standard revenue management technique to optimize over quantiles rather than prices \cite{TvR05}.

\subsubsection{Allocation vectorization.}

Even in a city of modest size, there are easily more than 10,000 locations to consider, which yields more than 100,000,000 allocation variables.
The authors are not aware of solvers that can solve such large problem instances in a matter of seconds, which is a requirement for our real-time application.
Luckily, because allocation neighborhoods are generally far smaller than the size of the entire region and there are operationally undesirable allocation pairs, $A^\text{max}$ and therefore any valid allocation is extremely sparse.
We vectorize the allocation matrix by constructing an allocation vector where each element corresponds to a valid allocation pair as indicated by a nonzero entry in $A^\text{max}$.

\subsubsection{Location pruning.}

We consider only locations that are affected by demand and supply changes to ensure solution efficiency.
Specifically, we only consider the union of two types of locations.
The first type comprises locations with nonzero demand, and the locations within their dispatch neighborhood.
The second type comprises locations with nonzero PPZ-assignable supply that have an allocation neighborhood that has any overlap with the dispatch neighborhood of locations with nonzero demand, and the locations within their allocation neighborhood.
These form our active set of locations that we consider in the mathematical program.
Combining location pruning with allocation vectorization, we are able to drastically reduce the number of decision variables, while provably preserving the optimal solution.
This reduces the number of variables to under 100,000, even for our largest markets.
Using the FICO Xpress quadratic program commercial solver on an AWS C5n instance featuring four 3.0 GHz Intel Xeon Platinum processors and 21 GiB of memory, we are able to consistently solve the problems in seconds and deliver fresh PPZ incentives every minute.



\section{Perturbation and Sensitivity Analysis}
\label{appendix-sensitivity}

In this appendix we provide analytical sensitivity results to demonstrate that the local sensitivity approach fails to accurately represent market conditions.

Consider Problem~\eqref{equation-driver-positioning-problem}, the driver positioning problem, except that drivers are fixed in their original locations and the only variable is the conversion quantile $y$.
Our problem is to maximize the expected sum of trip fares by setting appropriate multiplicative price modifiers $x \in \reals_+^n$, which affect the conversion of rider app-opens into requests $y$.
Solving this problem thus gives us the maximum revenue or bookings we can achieve from the market given the supply distribution that we currently have.
This simplified market-optimizing problem is
\begin{equation}
\begin{array}{ll}
\mbox{minimize}   & f^\text{obj} \left(x\right) = -f^T \left( d \circ y \right) \\
\mbox{subject to} & M \left(d \circ y\right) \preceq Ms_0 - r \text{.}
\end{array}
\label{equation-mop}
\end{equation}

Suppose Problem~\eqref{equation-mop} is feasible.
Because we have a linear program, strong duality holds.
Here, the optimal dual variables provide insights on the sensitivity of the optimal value with respect to the perturbations of the constraints.
In particular, they tell us how the optimal market revenue varies with respect to changes in the market-balancing conditions and, in turn, how demand, pricing, and supply may influence the revenue.

\subsection{The Perturbed Problem}
\label{subappendix-perturbed-problem}

We consider the following perturbed version of the original Problem~\eqref{equation-mop}:
\begin{equation}
\begin{array}{ll}
\mbox{minimize}   & f^\text{obj}\left(y\right) = -f^T \left( d \circ y \right) \\
\mbox{subject to} & M \left( d \circ y \right ) - \left(M s_0 - r \right) \preceq q
\end{array}
\label{equation-pmop}
\end{equation}
with variables $y$.
The problem coincides with the original optimization problem when $q=0$.
When $q>0$, we have relaxed the constraint; when $q<0$, we have tightened it.
Thus, the perturbed problem results from the original problem by tightening or relaxing the right side of the adjusted market-balancing inequality by $q$.

For the sake of clear exposition in this section, we will rewrite the optimization Problem~\eqref{equation-pmop} as
\begin{equation}
\begin{array}{ll}
\mbox{minimize}   & f^\text{obj}\left(y\right) \\
\mbox{subject to} & f^\text{mkt}\left(y\right) \preceq q \text{,}
\end{array}
\label{equation-pmop-simple}
\end{equation}
where $f^\text{mkt}$ is the left side of the perturbed problem's adjusted market-balancing condition.

We define $p^\star\left(q\right)$ as the optimal value of the perturbed Problem~\eqref{equation-pmop-simple}
\begin{equation}
p^\star\left(q\right) = \inf\left\{ f^\text{obj}\left(y\right) \;\middle|\; y\in\mathcal{D}, f^\text{mkt}\left(y\right) \preceq q \right\}\text{,}
\label{equation-opt-val-perturbed}
\end{equation}
where $\mathcal{D}$ is the domain of the optimization problem (\ie, the set of the points that are defined on the objective and constraint functions).
We can have $p^\star\left(q\right) = \infty$ correspond to the perturbations of the constraints that results in infeasibility.
Note that $p^\star\left(0\right) = p^\star$, the optimal value of the unperturbed problem.
Roughly speaking, the function $p^\star: \reals^{n} \rightarrow \reals$ gives the optimal value of the problem as a function of perturbations to the right sides of the market-balancing constraint [Constraint~\eqref{eqn-model-balance}].

The function $p^\star$ is a convex function of $q$.
To see this, consider the function
\[
P\left(y, q\right) = \left\{
\begin{array}{cl}
f^\text{obj}\left(y\right) & \quad f^\text{mkt}\left(y\right) \preceq q \\
\infty & \quad \text{otherwise.}
\end{array}
\right.
\]
Notice that $P$ is convex on its domain $\dom P = \left\{ \left(y, q\right) \;\middle|\; y\in\mathcal{D}, f^\text{mkt}\left(y\right) \preceq q \right\}$, which in turn is also convex.
We thus see that $P$ is convex jointly in $y$ and $q$, which means that $p^\star\left(q\right) = \inf_y P\left(y, q\right)$ is convex.

\subsection{Local Sensitivity Analysis}
\label{subappendix-local-sensitivity}

Let $\left(\lambda^\star, \nu^\star\right)$ be optimal for the dual of the unperturbed optimization Problem~\eqref{equation-mop}, where $\lambda$ is the dual variable that corresponds to the market-balancing constraint and $\nu$ is the dual variable that corresponds to the price modifier constraint.
Suppose now that $p^\star\left(q\right)$ is differentiable at $q=0$.
Then, provided that strong duality holds, the optimal dual variable $\lambda^\star$ equals the gradient of $p^\star$ at $q=0$ with respect to $q_i$, that is,
\begin{equation}
\lambda_i^\star = -\frac{\partial p^\star\left(0\right)}{\partial q_i}\text{.}
\label{equation-local-sensitivity}
\end{equation}
This means that if the above conditions hold, the optimal Lagrange multipliers are exactly the local sensitivities of the optimal value with respect to constraint perturbations.
In economics, $\lambda$ and $\nu$ are often referred to as the natural or equilibrium prices or ``shadow prices'' of their corresponding constraints.
In the case of the market-balancing conditions, these can be thought of as the price that we should accord to (marginal) supply in the dispatch neighborhoods.
That is, $\lambda_i^\star$ is the marginal value of supply in PPZ's context.

Locally, this interpretation is symmetric: decreasing the supply count in the $i$th location's dispatch neighborhood by a small amount (\ie, taking $q_i$ small and negative) yields a change in the optimal market revenue objective value $-p^\star$ of approximately $-\lambda_i^\star q_i$; increasing the supply count in the $i$th location's dispatch neighborhood by a small amount (\ie, taking $q_i$ small and positive) yields an increase of approximately $\lambda_i^\star q_i$.
Thus, Equation~\eqref{equation-local-sensitivity} gives us precisely the marginal value of supply to use in a local sensitivity approach.

This local sensitivity result gives us a quantitative measure of how active a market-balancing constraint is at the optimum $y^\star$.
If $f_i^\text{mkt} < 0$, then the constraint is inactive, and it follows that the supply count can be decreased or increased a little without affecting the optimal market revenue objective value.
By complementary slackness, the associated Lagrange multiplier $\lambda_i^\star$ must be zero.
However, now suppose that $f_i^\text{mkt} = 0$, that is, the market-balancing constraint for the $i$th location's dispatch neighborhood is active at the optimum.
The $i$th optimal Lagrange multiplier tells us how active the constraint is: if $\lambda_i^\star$ is small, it means that the supply count can be decreased or increased a little without much effect on the optimal market revenue objective value; if $\lambda_i^\star$ is large, it means that if the supply count is decreased or increased a little a bit, the effect on the optimal market revenue objective value will be great.

This analysis motivates a natural approach to allocate PPZs based on $\lambda^\star$: PPZs should incentivize drivers to reposition from locations $i$ with small $\lambda_i^\star$ to locations $j$ with large $\lambda_j^\star$.
Based on the above, one would expect this to increase the market revenue.
Unfortunately, as we will demonstrate, the local sensitivity analysis does not generalize beyond the point at which the marginal value of supply was evaluated.

\subsection{Global Sensitivity Analysis}
\label{subappendix-global-sensitivity}

To examine how the local sensitivity results break down, consider Equation~\eqref{equation-opt-val-perturbed}.
For all $q$, we will show that
\begin{equation}
p^\star\left(q\right) \geq p^\star\left(0\right) - \lambda^{\star T}q\text{.}
\label{eqn-global-sensitivity}
\end{equation}
To establish this inequality, suppose $y$ is feasible for the perturbed problem.
Then we have, by strong duality,
\begin{align*}
p^\star\left(0\right) = g\left(\lambda^\star, \nu^\star\right)
&\leq f^\text{obj}\left(y\right) + \lambda^{\star T} f^\text{mkt}\left(y\right) + \nu^{\star T}f^\text{bal}\left(y\right) \\
&\leq f^\text{obj}\left(y\right) + \lambda^{\star T} q + \nu^{\star T} 0 \\
&= f^\text{obj}\left(y\right) + \lambda^{\star T} q \text{.}
\end{align*}
Here, $g$ is the Lagrange dual function and the first inequality follows from its definition.
Recall also that $\lambda^\star \succeq 0$ by definition of the Lagrange dual problem.
We thus have
\begin{align*}
f^\text{obj}\left(y\right) \geq p^\star\left(0\right) - \lambda^{\star T} q \text{,}
\end{align*}
which leads to Inequality~\eqref{eqn-global-sensitivity}.

Recall that $p^\star\left(q\right)$ is the negative of the perturbed optimal market revenue objective value.
For clarity, we can rewrite the inequality constraint as
\begin{align*}
-p^\star\left(q\right) \leq \lambda^{\star T} q - p^\star\left(0\right)
\end{align*}
and make the two following observations:
\begin{enumerate}
\item
Suppose $\lambda_i^\star$ is large and we decrease the supply count in the $i$th location's dispatch neighborhood; that is, we tighten the $i$th constraint and choose $q_i < 0$.
Then the optimal market revenue objective value of the objective $-p^\star\left(q\right)$ will decrease greatly.
\item
Suppose $\lambda_i^\star$ is small and we increase the supply count in the $i$th location's dispatch neighborhood; that is, we loosen the $i$th constraint and choose $q_i > 0$.
Then the optimal market revenue objective value of the objective $-p^\star\left(q\right)$ will not increase too much.
\end{enumerate}

So far, these observations align with the results obtained from the local sensitivity analysis. The inequality established and the two observations above provide an upper bound on $-p^\star\left(q\right)$, the optimal market revenue objective value of the perturbed version of the original market-optimizing problem.
Notice, however, that it establishes no lower bound of the perturbed optimal market revenue objective.
We thus see that the results are not symmetric with respect to increasing or decreasing the supply count in the market-balancing constraint.

To illustrate this more clearly, consider the case where $\lambda_i$ is large and we slightly increase the supply count in the $i$th location's dispatch neighborhood; that is, loosen the $i$th constraint slightly and choose $q_i > 0$ for some small $q_i$.
In this case, our inequality is not helpful in establishing any conclusion about how the perturbed optimal market revenue objective changes; it certainly does not imply that the perturbed optimal market revenue objective $-p^\star\left(q\right)$ increases considerably.
In other words, \emph{a large optimal dual variable value for some location's dispatch neighborhood does not imply that adding more drivers to it will improve the market conditions}.

Recall the rule of incentivizing drivers to move from locations $i$ with small $\lambda_i^\star$ to locations j with large $ \lambda_j^\star$ established from our local sensitivity results.
Unlike what our local sensitivity results suggest, our global sensitivity result tells us that a large ``shadow price'' or ``marginal value of supply'' for some location does not imply that incentivizing drivers to resposition to that location will yield market revenue gains.
Given that incentivizing drivers to reposition away from a location~$i$ may, even with small~$\lambda_i^\star$, yield large market revenue drops, simply using a static value of supply defined by an optimal Lagrange multiplier can hurt us because of this asymmetry.
Recalling \Cref{Figure3}, this asymmetry is apparent in both the analytical and the empirical estimates.
Thus, a static set of shadow prices is a poor way to coordinate how supply is managed.
If at all useful, it might be for visualizing snapshots of and providing spatial intuition for how severe the supply shortage is across a city.
To actually coordinate supply management tools, we needed a fundamentally different approach, as we have discussed in this paper.

\section{Video of PPZs Served in Production}
\label{appendix-video}

\begin{figure}[t]
    \centering
    \includegraphics[width=0.8\linewidth]{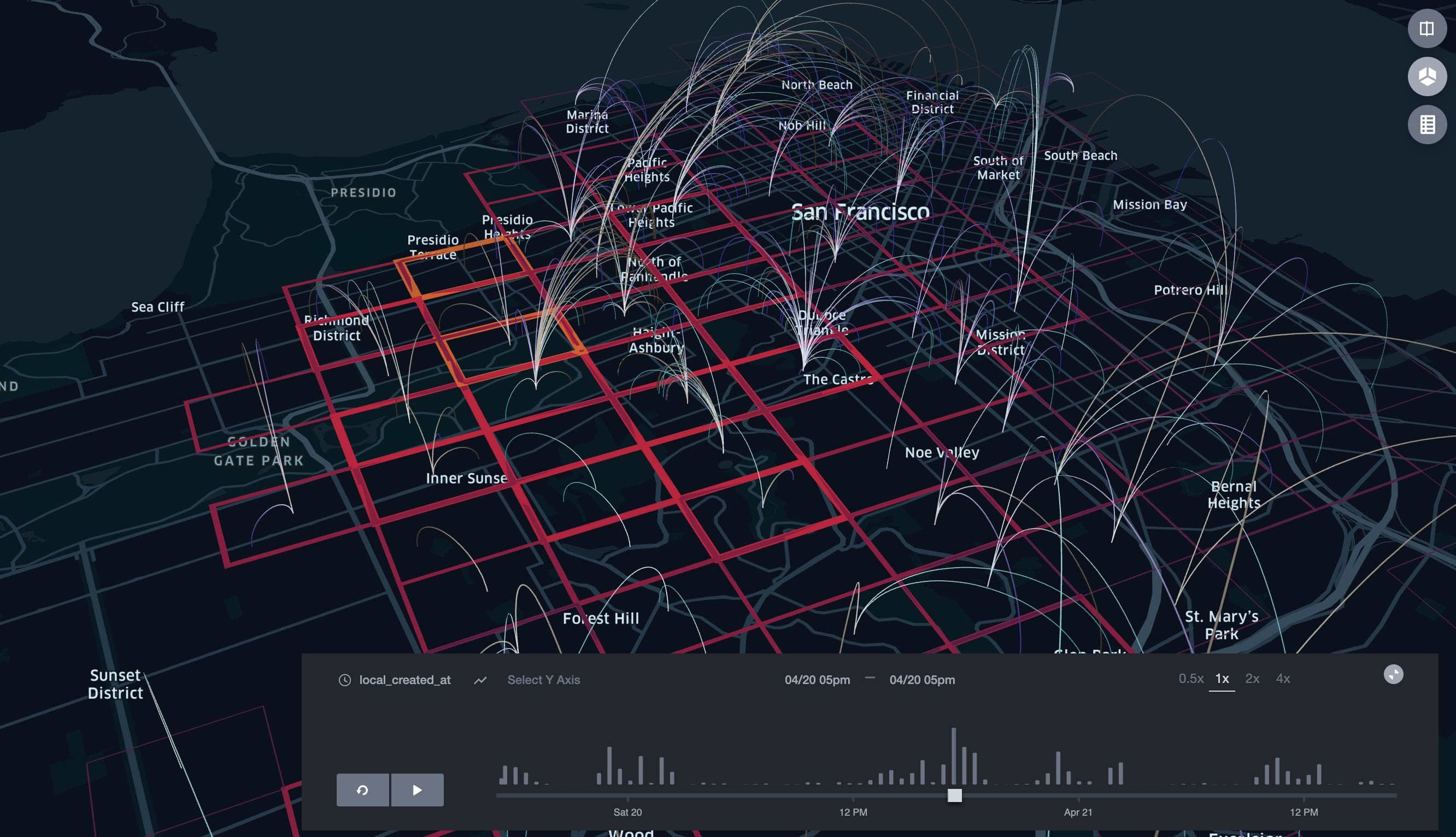}
    \caption{This snapshot of the video illustrates PPZs actually served in the San Francisco Bay Area. Note that the arc colors are used to indicate whether the drivers complied with and earned the incentives.}
    \label{Figure9}
\end{figure}

An videographic illustration of actual PPZ incentives being served in production for the San Francisco Bay Area can be found at \url{https://youtu.be/mlxY-qxlb6w}.
The data for this video were obtained from the period of April 20--21, 2019.
The video plays back the PPZs created for and earned by actual drivers and provides some intuition about when and where repositioning opportunities are.
\Cref{Figure9} is a snapshot of the video, with the colored arcs representing the origin and destination of the PPZ directions.
The PPZ origin is represented by the end of the arc in a lighter hue and the destination the one in a darker hue.
The purple arcs indicate PPZs that were eventually satisfied and earned by drivers, whereas the teal arcs indicate PPZs that were not satisfied.
The highlighted boxes indicate the location buckets used in our algorithm, with red indicating areas of higher actual PT and yellow areas of lower PT.
When there is no PT, there is no highlight.
The rate of PPZs generated is indicated in a time-series plot at the bottom of the video, providing a sense of the seasonality of when the best opportunities to allocate PPZs are (typically, morning and evening commutes to and from the residential and downtown areas of the city).

\end{document}